# Production of the artificial $^{51}$Cr neutrino source in the BEST project

**S.N. Danshin[a], V.N. Gavrin[a], V.V. Gorbachev[a], T.V. Ibragimova[a], B.A. Komarov[a], J.P. Kozlova[a], A.A. Martynov[a], E.P. Veretenkin[a], L.V. Akimov[b], A.V. Kupriyanov[b], A.P. Malkov[b], A.L. Petelin[b], V.V. Pimenov[b], E.G. Romanov[b], S.A. Sazontov[b], E.M. Tabakin[b], V.A. Tarasov[b], I.V. Torgashov[b], V.A. Uzikov[b], A.I. Zvir[b] and A.A. Kalaschnikova[c]**

[a] *Institute for Nuclear Research of the Russian Academy of Sciences, Prospect 60-letiya Oktyabrya, 7a, Moscow, 117312, Russian Federation*

[b] *Research Institute of Atomic Reactors, Zapadnoye Shosse, 9, Ulyanovsk region, Dimitrovgrad, 433510, Russian Federation*

[c] *Bauman Moscow State technical University, Baumanskay 2 st., 5, Moscow, 105005, Russian Federation*

E-mail: julia@inr.ru

ABSTRACT:

The production of the artificial $^{51}$Cr neutrino source with activity > 3 MCi for the experiment BEST is presented. This procedure consisted of making a $^{50}$Cr target and irradiating it with thermal neutrons in a nuclear reactor SM-3. The production of a target in the form of disks with a thickness of 4 mm and a diameter of 84 and 88 mm included enrichment (to 96.5% in $^{50}$Cr) of natural chromium in the form of oxyfluoride by gas centrifugation, electrolytic reduction and refining of metallic chromium, as well as the formation of chromium disks by spark plasma sintering. Simulations of various source geometries, neutron flux and nuclear transmutation were carried out to validate the design of the source, the irradiation device and the transport container, the required chemical purity of the target and the irradiation schedule in the reactor. The calculated activity of the source after 75 effective days of irradiation was 3.55 MCi. The activity of the source was measured by the calorimetric method and amounted to 3.41 MCi at the time of its delivery to the Baksan Neutrino Observatory. This is the most intense chemically pure neutrino source ever produced.

## 1. Introduction

In studies of the composition of the modern Universe, one of the most important tasks at present is to obtain complete information about the neutrino sector. One of the priority tasks in this direction is to test the hypothesis of the existence of sterile neutrinos - hypothetical particles that interact with ordinary matter exclusively in a gravitational way and, according to modern ideas, may or might form part of dark matter. First evidence for the possibility of the existence of such neutrinos– was indicated by the results of accelerator experiments (LSND [1] and MiniBooNE [2]), experiments on calibration of SAGE and GALLEX detectors with artificial neutrino sources [3,4], commonly referred to as gallium anomaly, as well as the revision of some reactor experiment results [5].

To search for sterile neutrinos, Baksan Experiment on Sterile Transitions (BEST) was proposed and conducted by studying the neutrino flux from a compact artificial high-intensity source based on the radionuclide $^{51}$Cr in a two-zone metallic gallium target of the gallium-germanium neutrino telescope (GGNT) in the Baksan Neutrino Observatory (BNO) [6]. GGNT is located below the earth's surface at a depth of 4800 m.w.e, where the main contribution to the background will be given by neutrinos from the Sun, the flux of which is well known from the long-term measurements. By comparing the neutrino capture rates in the inner and outer zones of gallium target, as well as the expected and measured capture rates, the search for short baseline neutrino oscillations was proposed.

The neutrino interaction rate was determined by the rate of neutrino capture by $^{71}$Ga nuclei with the formation of $^{71}$Ge atoms, which were extracted and counted independently from each zone. The rates at each distance were found to be similar, but 20-24% lower than expected, thus reaffirming the gallium anomaly. These results are consistent with $\nu_e \rightarrow \nu_s$ oscillations with a relatively large $\Delta m^2$ (>1 eV$^2$) and mixing $\sin^2 2\theta$ ($\approx 0.4$) [7].

It was assumed that the neutrino source based on $^{51}$Cr would have an activity of the order of 3 MCi, which made it possible to achieve a sensitivity of several percent to the disappearance of electron neutrinos. An artificial $^{51}$Cr neutrino source can be obtained by irradiating a target containing $^{50}$Cr with a neutron flux by the $^{50}$Cr(n, γ)$^{51}$Cr reaction in the SM-3 reactor of the Joint Stock Company "State Scientific Center Research Institute of Atomic Reactor" (JSC "SSC RIAR"). Considering the time required for the assembly and delivery of the source, the activity of the $^{51}$Cr source at the end of the irradiation should be at least 3.5 MCi.

The overall dimensions of the source with biological protection and the external housing with the grip handle (see section 5) are limited by the design of the two-zone gallium target in the BEST experiment: diameter is no more than 160 mm, height is no more than 230 mm.

Such a high-active compact source can be produced by irradiating chromium with a high enrichment of the $^{50}$Cr isotope (the $^{50}$Cr content in natural chromium is 4.35%). Due to the high cost of enrichment, it is possible to produce a relatively small (up to 4.5 kg) amount of enriched chromium. The preliminary estimates have shown that the required values of the specific activity of $^{51}$Cr (about 1 kCi/g) can be achieved only in the most high-flux region of the SM-3 reactor: the central neutron trap [8].

## 2. The type of reactor target for $^{50}$Cr irradiation
### 2.1 Neutron-physical simulations

The variant neutron-physical simulations and calculations of nuclear transmutation were carried out to substantiate the design of the irradiation device (IRD), which allows to obtain the activity ~ 3.5 MCi $^{51}$Cr at the end of bombardment. Calculations of neutron-physical characteristics (group density of the neutron flux, neutron gas temperature, energy release) in the volume of irradiated chromium items were carried out using the MCNP program [9], which used the Monte Carlo method for modeling the transport of particles (neutrons, electrons, photons) in three-dimensional geometry.

For numerical simulation of the accumulation of $^{51}$Cr and other radionuclides of interest, the ChainSolver program from the ORIP_XXI software package developed at the JSC «SSC RIAR» was used [10]. The program is designed to calculate the transmutation of nuclei in samples irradiated in the neutron field of a nuclear reactor. The program algorithm allows to consider the self-shielding of the resonances of the cross sections of neutron capture by nuclei, the depression of the thermal neutron flux in the material, as well as the real irradiation schedule (the schedule of the reactor operation and the rearrangement of the irradiated ampoule in the position with different neutron fluxes and spectra).

Initial simulations considered irradiation of rods of various number, shape, and size, made of metallic chromium enriched in $^{50}$Cr. [11]. The calculations shown that none of the variants allows obtaining a source with an activity of ~3.5 MCi. A common disadvantage of all the rod structures of the irradiation devices is their limited capacity (about 3 kg of metal chromium), as well as many rods, from which it was necessary to assemble the source quickly.

Next, an irradiation device with a chrome target in the form of disks was considered. The disks size was limited by the diameter of the central neutron trap of the reactor SM-3 (110 mm). Considering the necessary technological gaps, the maximum diameter of the disks was limited to 90 mm.

Six variants of disks size were considered:
1. 18 disks Ø90 mm, 3.8 mm thick with a central hole Ø4 mm. The total weight of chromium is 3000 g.
2. 36 disks Ø90 mm, 1.9 mm thick with a central hole Ø4 mm. The total weight of chromium is 3000 g.
3. 20 disks Ø90 mm, 3.8 mm thick with a central hole Ø4 mm. The total weight of chromium is 3334 g.
4. 20 disks Ø88 mm, 4 mm thick with a central hole Ø4 mm. The total weight of chromium is 3355 g.
5. 10 disks Ø88 mm, 4 mm thick with a central hole Ø22 mm and 10 disks Ø84 mm, 4 mm thick with a central hole Ø4 mm. The total weight of chromium is 3104 g. In the IRD, disks of different sizes are installed alternately to organize the leaking of the coolant.

6. 13 disks Ø88 mm, 4 mm thick with a central hole Ø22 mm and 13 disks Ø84 mm, 4 mm thick with a central hole Ø6 mm. The disks are evenly distributed along the height of the core (from -17.5 cm to 17.5 cm relative to the central plane of the core). The total weight of chromium is 4035 g.

Table 1 shows the neutron-physical characteristics averaged over the disks and the activity of $^{51}$Cr at the time of the irradiation end, the operating mode of the reactor was assumed to be the same as in the case of a rod irradiation device.

Table 1. Results of neutron-physical calculations of disk IRD.

| Option no. | Neutron flux at 100 MW power of reactor, cm$^{-2}$·s$^{-1}$ | | | Neutron gas temperature, K | Activity $^{51}$Cr, MCi |
|---|---|---|---|---|---|
| | 0 – 0.5 eV | 0.5 eV – 0.1 MeV (Per unit lethargy*) | 0.1 – 20 MeV | | |
| 1 | 4.47E+14 | 9.82E+13 | 9.79E+14 | 686 | 3.13 |
| 2 | 4.58E+14 | 9.99E+13 | 9.67E+14 | 676 | 3.24 |
| 3 | 4.34E+14 | 9.62E+13 | 9.55E+14 | 686 | 3.37 |
| 4 | 4.26E+14 | 9.58E+13 | 9.49E+14 | 692 | 3.32 |
| 5 | 4.46E+14 | 9.67E+13 | 9.46E+14 | 685 | 3.23 |
| 6 | 3.96E+14 | 9.07E+13 | 8.87E+14 | 695 | 3.62 |

*Lethargy unit defined as $u = \ln(\frac{E_{up}}{E_{low}})$, where in this case $E_{up}$=0.1 MeV and $E_{low}$=0.5 eV are upper and lower energy limits of intermediate neutron spectrum. It is convenient to use flux per unit lethargy (Φ/u) in calculation of reaction rates (radiative capture, fission, etc.), because in the neutron slowing-down region Φ~1/E, so the ratio Φ/u is a constant at any $E_{up}$ and $E_{low}$ within this region [12]

As follows from Table 1, the use of the disks (option 6) allows to bring the activity $^{51}$Cr to the required values, primarily because of the possibility to increase the mass of the starting chromium.

**2.2. Design of the irradiation device**

The following IRD design was proposed for the irradiation of $^{50}$Cr and the accumulation of $^{51}$Cr in the amount of ~ 3.5 MCi (Figure 1).

The IRD consists of a housing, a central support rod on which spacer elements, upper and lower grids are fixed. The housing, the central rod and the spacing elements are made of zirconium alloy, the upper and lower grids are made of stainless steel. 26 disks of two standard sizes are used for irradiation: a disk with an external diameter of 88 mm, an internal diameter of 22 mm and a thickness of 4 mm, a disk with an external diameter of 84 mm, an internal diameter of 6 mm and a thickness of 4 mm. In the IRD, the disks are placed alternately with each other, which ensures the creation of a circulation path between the disks and their effective cooling by the reactor coolant. The height spacing of the disks is carried out using zirconium elements, which consist of a perforated disk substrate and spacers placed at an angle of 120° to each other. The IRD is installed in a neutron trap instead of a separator made of zirconium pipes.

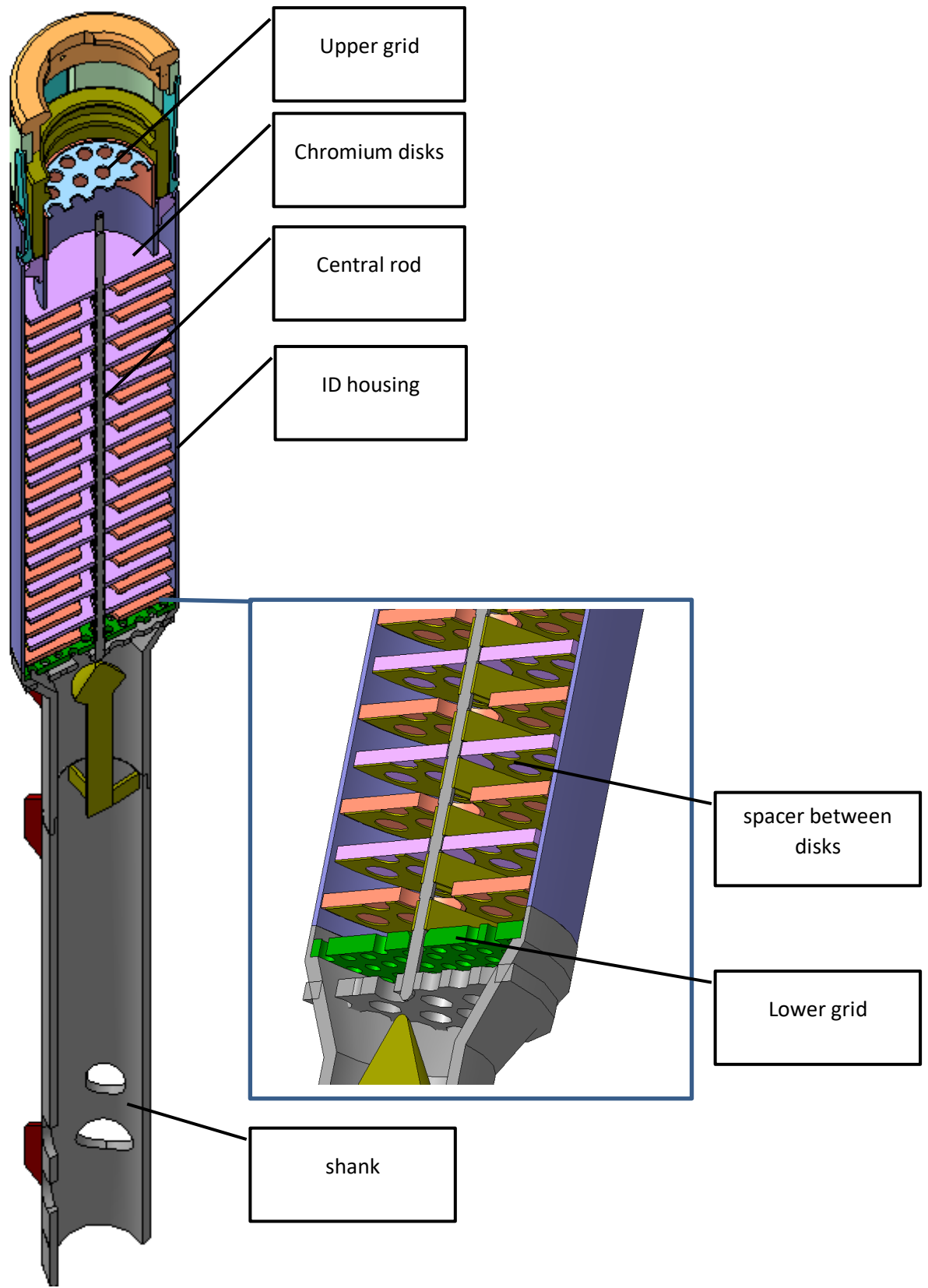

Figure 1. Design of the irradiation device for the $^{51}$Cr accumulation.

Since IRD had the direct contact with the coolant, test experiments were carried out before the irradiation. It was shown that the contamination of chromium by radionuclides from the coolant was negligible.

## 2.3 Calculation of nuclear transmutation and requirements for the chemical composition of chromium

An important characteristic of the source, showing the possibility of safe operation with it, is the exposure dose rate (EDR). The EDR is determined both by the radiation of the $^{51}$Cr and of the radionuclides formed in the reactor during the irradiation of the target.

Based on the preliminary results, the values of achievable levels of the most common activated impurities were obtained. These values were used to calculate the EDR and develop recommendations for reducing the content of critical impurities if necessary (Table 2). The weight of chromium 3860 g is determined based on the sizes of disks ⌀88×22 mm and ⌀84×6 mm with thickness 4 mm (13 pieces of each nominal value) and the density of chromium 6.62 g/cm$^3$. The $^{50}$Cr isotope enrichment was assumed to be 98%.

Table 2. The mass of chromium and impurity elements for calculations of nuclear transmutation.

| Element | Cr | Na | Mn | Fe | Zn | Sb |
|---|---|---|---|---|---|---|
| Mass, g (content, ppm) | 3860 | 0.04 (10) | 0.004 (1) | 0.4 (100) | 0.008 (2) | 0.02 (5) |

The EDR was calculated using the MCNP program, which simulates the transport of photons and electrons in the substance by the Monte Carlo method [9]. In addition to the $^{51}$Cr, the contribution of the following activation products of impurity elements was considered: $^{24}$Na, $^{56}$Mn, $^{59}$Fe, $^{60}$Co, $^{65}$Zn, $^{124}$Sb. It should be noted that the contribution of activation products of titanium, tin and barium to the EDR was not taken into account due to the low yield during irradiation ($^{46}$Sc, $^{113}$Sn, $^{133}$Ba, respectively), copper due to the short half-life $^{64}$Cu (12.7 hours), nickel due to the low radiation energy (17 keV, $^{63}$Ni - beta emitter). $^{60}$Co was produced by the Fe activation. Mass-spectrometric analysis of chromium showed an Ag impurity content below the detection limits of 0.1 ppm. Therefore, the possibility of chromium contamination by $^{110m}$Ag was not considered.

The results of calculating the accumulation of chromium isotopes and activation products of impurities that significantly affect the dosimetric situation near the source at the end of irradiation are shown in Table 3.

Table 3. The calculation of the nuclear transmutation.

| Activity, Ci | | | | | | | Mass, g | | | |
|---|---|---|---|---|---|---|---|---|---|---|
| $^{51}$Cr | $^{24}$Na | $^{56}$Mn | $^{59}$Fe | $^{60}$Co | $^{65}$Zn | $^{124}$Sb | $^{50}$Cr | $^{51}$Cr | $^{52}$Cr | $^{51}$V |
| 3.64E06 | 3.78 | 4.48 | 0.086 | 0.022 | 0.058 | 7.25 | 3677.7 | 39.19 | 0.48 | 67.23 |

The radiation situation during the operation of a neutrino source in the BNO is determined by the value of the EDR at a meter from the side surface of the biological protection. Table 4 shows the values of the relative contributions to the EDR of radionuclides formed during the activation of impurities.

Table 4. Relative contribution of radionuclides to the EDR at one meter from the side surface of the source biological protection shield.

| Radionuclide, $T_{1/2}$ | Major radiations, γ-rays, keV | Contribution to the total EDR, % (0,1,2,3,4,5 days after irradiation) | | | | | |
|---|---|---|---|---|---|---|---|
| | | 0 | 1 | 2 | 3 | 4 | 5 |
| $^{24}$Na, 14.997 h | 1368.6 2754.0 | 54.8 | 37.9 | 16.9 | 6.3 | 2.2 | 0.7 |
| $^{56}$Mn, 2.58 h | 846.8 1810.7 | 15.4 | 0.1 | 0.0 | 0.0 | 0.0 | 0.0 |
| $^{59}$Fe, 44.495 d | 1099.25 1291.59 | 0.2 | 0.4 | 0.5 | 0.6 | 0.6 | 0.6 |
| $^{60}$Co, 5.27 y | 1332.5 1173.2 | 0.1 | 0.2 | 0.3 | 0.4 | 0.4 | 0.4 |
| $^{65}$Zn, 243.93 d | 1115.54 8.05 | 0.1 | 0.1 | 0.1 | 0.2 | 0.2 | 0.2 |
| $^{124}$Sb, 60.2 d | 602.7 1990.97 | 27.3 | 56.7 | 76.0 | 85.8 | 89.7 | 91.1 |
| $^{51}$Cr, 27.7 d | 4.95 320.08 | 2.2 | 4.6 | 6.2 | 6.8 | 7.0 | 7.0 |

The data in Table 4 show that, with the chemical composition of the target taken for calculations, at the time of arrival of the neutrino source at the experiment site (approximately two days after the end of irradiation) the EDR would be determined by the isotopes $^{124}$Sb and $^{24}$Na. The absolute value of the EDR will be 2500 mSv/hour, of which only ~170 mSv/hour is due to the radiation of the $^{51}$Cr itself.

Thus, the content of antimony and sodium in the starting material should be about one and a half orders of magnitude lower compared to the calculated model.

## 3. A metal $^{50}$Cr preparation

### 3.1 $CrO_2F_2$ enrichment in $^{50}$Cr

To obtain the source activity required for the experiment BEST, more than 3 MCi, the reactor target must be made of chromium with a $^{50}$Cr isotope content of ≥ 97%. The content of the $^{50}$Cr isotope in the natural chromium is 4.35%. At the first stage, the possibility of using enriched chromium from the source of the SAGE calibration experiment was considered [3]. However, the complexity of transporting radioactive material from another country (Kazakhstan) and the need to recycle material with residual $^{60}$Co led to the rejection of this idea. To obtain the enriched $^{50}$Cr, the gas centrifugation method was chosen as the most economical and high-performance method of isotope separation. The chromium compound for gas centrifugation must meet the following requirements: first, it must be in a gaseous state within the operating temperature range of gas centrifuges, and second, it must contain only monoisotopic elements, except for chromium. Such a compound is chromyl fluoride $CrF_2O_2$, which is characterized by: high saturated vapor pressure of 760 mm Hg at a temperature of 29.6°C [13]; chemical resistance to the materials used in the equipment and chemical stability at elevated temperatures, which prevents the formation of non-volatile chromium compounds during centrifugation. Oxygen and fluorine in $CrF_2O_2$ have one stable isotope each.

At the first stage, 380 kg of chromyl fluoride was synthesized for the enrichment. For the synthesis, the most technologically advanced process of direct fluorination of $CrO_3$ with molecular fluorine under flow conditions was selected. The synthesis was carried out at a temperature of 260°C and a pressure of 140-150 mmHg by the reaction:

$$2CrO_3 + 2F_2 \rightarrow 2CrO_2F_2 + O_2.$$

Chromium enrichment was carried out at the Electrochemical Plant JSC using cascades of gas centrifuges in two stages: at the first stage, 25% enrichment was achieved, and at the second stage – 98% enrichment. In addition to isotope separation the gas centrifugation technology provided additional purification of $CrO_2F_2$ from impurities. The chromyl fluoride was then hydrolyzed to chromium anhydride $CrO_3$ by the reaction:

$$CrO_2F_2 + H_2O \rightarrow CrO_3 + 2HF.$$

As a result of the enrichment and hydrolysis of chromyl fluoride, 8875 g of chromium anhydride containing 4500 g of chromium was obtained. The chromium was enriched to 98.46% in $^{50}Cr$.

**3.2 Electrolytic reduction and refining of metallic enriched chromium.**

The next step in the production of the reactor target was the electrolytic reduction to chromium metal. For this purpose, an electrolyser with a capacity of 25 g/hour was developed and manufactured. The scheme of the electrolyzer with the main dimensions is shown in Figure 2.

The electrolyzer is an elliptical titanium vessel with a titanium water-cooled heat exchanger. In the vessel there is a cathode made of a rectangular sheet of stainless steel and, on both sides of it, two anodes made of lead sheet. The working volume of the electrolyzer is 15 liters, the working area of the cathode is 600 cm². Metallic chromium was released at the cathode.

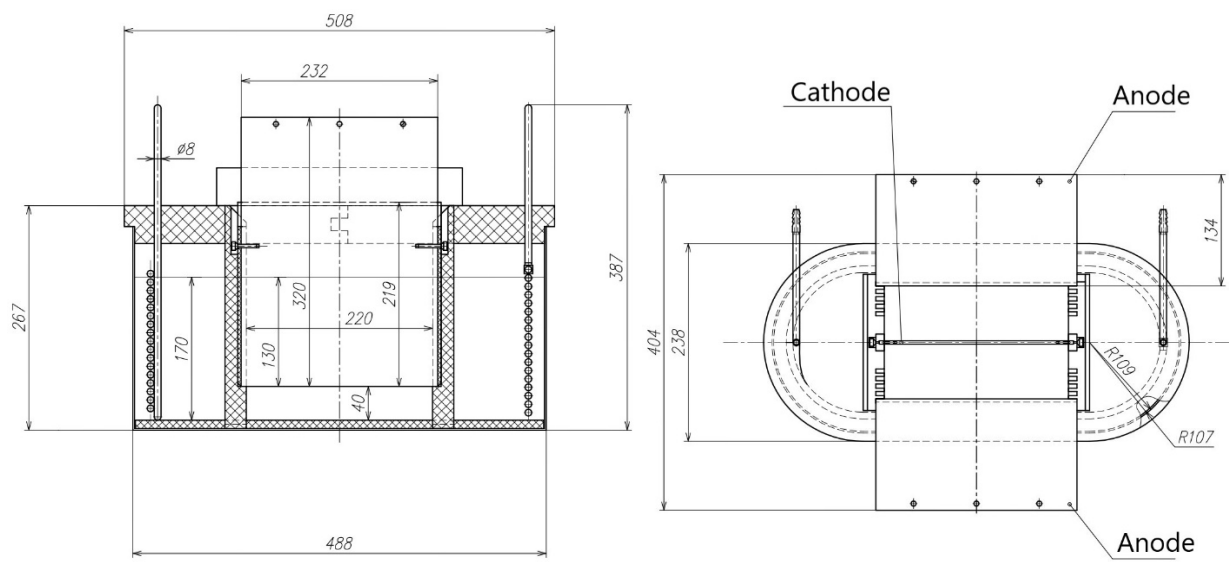

Figure 2. The scheme of the electrolyzer, dimensions are given in mm.

The electrolysis was carried out from an electrolyte with an initial concentration of chromium anhydride $CrO_3$ 300 g/l and sulfuric acid 3 g/l. Then 10 g/l of oxalic acid was added to the electrolyte once for the primary production of trivalent chromium. During the electrolysis process, the electrolyte composition was periodically checked, and the chromium anhydride content was maintained in the range of 200-330 g/l by adding dry chromium anhydride. The volume of the electrolyte is kept constant by adding water to compensate for its losses due to decomposition during electrolysis and evaporation.

At a current I = 180-200 A, voltage U = 4.5-6 V and temperature T = 18-23 ° C, the average efficiency of the electrolyzer was 20-25 g/hour for metallic chromium.

Due to a sharp decrease in the current output of chromium with a decrease in the concentration of chromium in the electrolyte, electrolysis can be carried out at a content of chromium anhydride of at least 180 g/l. When this value was reached, chromium was concentrated by evaporation of the electrolyte with simultaneous correction of the sulfuric acid content by removing its excess by adding barium hydroxide $Ba(OH)_2$. When the volume of the electrolyte decreased below 15 l, the volume of the cell was reduced to ~5 l by using special blocks and inserts from fluoropolymer. In this case, the electrolysis was carried out with a single anode at a current I = 90 – 100 A with an efficiency of 10-12 g/hour. The final stage of electrolysis was carried out on a laboratory electrolyzer with a volume of 0.8 liters and an efficiency of 4 g/hour.

As a result of electrolysis, 4215 g of electrolytic metallic chromium flakes were obtained from 8875 g of enriched chromium anhydride. The electrolytic chromium contained a significant amount of oxygen (~ 4400 ppm) and lead (~ 3 ppm) from Pb anode. The most effective method of refining electrolytic chromium is high temperature annealing of metallic chromium flakes in a hydrogen stream. The refining of metal chromium with hydrogen was carried out in JSC "POLEMA" in the following technological mode: the annealing temperature was 1540 °C, the annealing time was 12 hours, the total consumption of hydrogen per cycle was not less than 350 m$^3$. As a result, 4149 g of refined enriched chromium was obtained. Since the refining took place in a hydrogen furnace, which is constantly used for the purification of industrial batches of chromium of natural isotopic composition, the isotope exchange occurred at high temperature. So, the enrichment of chromium decreased slightly to 96.5% in the isotope $^{50}$Cr. After refining, the oxygen content in the chromium was 30 ppm and the lead content was < 0.05 ppm.

### 3.3 Manufacturing of a metallic chromium disks.

The metal chromium powder was obtained by grinding refined chromium flakes using a The Mortar Grinder RM 200 with an agate grinding tools. The use of agate was due to its high hardness, which significantly reduced the penetration of structural materials into the final product. In addition, the agate consists of 99.9% SiO$_2$, so the penetration of a small amount of this compound is not dangerous for the radiation purity of the neutrino source. The chromium flakes were ground to a powder with a particle size of no more than 315 microns.

The chemical composition of chromium powder was analyzed by the ICP-MS method with the sample dissolution in JSC "POLEMA", National Research Center "Kurchatov Institute" – IREA, non-metallic impurities were analyzed by infrared absorption in JSC "POLEMA". The contents of the critical elements are presented in Table 5. These results meet the requirements for the sintering powder (non-metallic impurities) and for a chromium target for thermal neutron irradiation in a nuclear reactor (metallic impurities). The content of critical impurity elements of sodium and antimony, the activation products of which determine the EDR, was only 5 and < 0.1 ppm, respectively.

Table 5. The content of the critical elements in chromium powder after annealing and grinding of the flakes.

| Element | Content, ppm | Critical content, ppm |
|---|---|---|
| C | 35 | |
| N | 20 | |
| O | 30 | |
| Na | < 5 | 0.3 |
| Mn | 0.1 | 1 |
| Fe | 90 | 100 |
| Co | <0.1 | 2 |
| Zn | < 0.5 | 5 |
| Sb | < 0.1 | 0.15 |

The technology of spark plasma sintering (IPS) was used to produce of metallic chromium disks. The IPS technology combines the methods of hot pressing and electrostimulated sintering. According to this technology, the compaction of the starting powder occurs under the pulsed currents and discharge plasma formed in the gaps between neighboring powder particles, which sharply stimulates diffusion sintering. Heating the substance by passing pulses of electric current allows to significantly reduce the temperature and the sintering time compared to conventional sintering and hot pressing. The obtained material has a high density and fine-grained structure. In addition, disks of the required geometry can be directly obtained by using the IPS technology with minimal material loss.

Powder sintering was carried out at the hybrid spark plasma sintering unit (KCE H-HP D 25-SD, FCT, Germany) at the STANKIN Moscow State Technical University. The process took place at a temperature of 1425°C and a pressure of 47.7 MPa. The specific density of the pressed chromium was 98.7% of the theoretical one, the porosity was 0%, and the bending strength was 170 MPa.

Using the developed technology [14], 13 disks with a diameter of 84 mm and 13 disks with a diameter of 88 mm and a thickness of 4 mm were made (Figure 3). The inner hole of small disks with a diameter of 6 mm was formed by electroerosive cutting; the inner hole of large disks (Ø 22 mm) was formed directly during their sintering without loss of material. The total weight of enriched chromium disks was 4007.5 g.

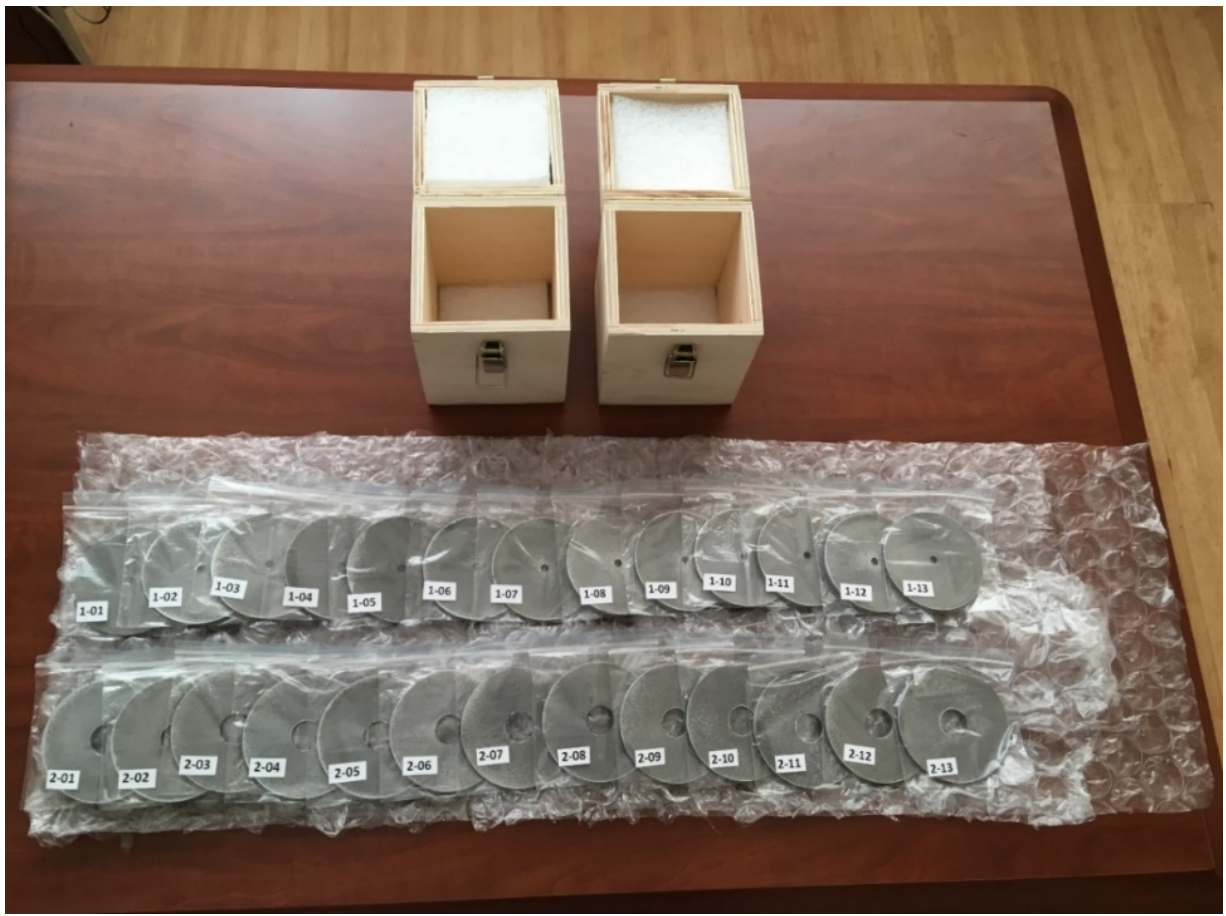

Figure 3. 26 disks of two standard sizes made of enriched chromium.

## 4. Irradiation of metal $^{50}$Cr targets in the SM-3 reactor

Loading a large amount of metallic chromium (about 4 kg) into the neutron trap significantly affects the neutron-physical characteristics of the reactor. The detailed experimental studies were carried out on a physical model of the SM-3 reactor.

To determine the energy release in the elements of the IRD numerical simulations of the neutron-physical characteristics were performed. The average neutron flux with an energy of 0-0.5 eV was $3.89\times10^{14}$ cm$^{-2}\cdot$s$^{-1}$, with an energy of 0.5 eV - 0.1 MeV was $9.67\times10^{13}$ cm$^{-2}\cdot$s$^{-1}$, with an energy of 0.1 - 20 MeV was $9.22\times10^{13}$ cm$^{-2}\cdot$s$^{-1}$, the average neutron gas temperature was 698 K, and the average energy release per disk was 28.9 W/g.

The thermohydraulic analysis of the stationary cooling mode of the IRD with chromium disks in SM-3 neutron trap was carried out using the Flow Simulation program of the SolidWorks package [15]. It was shown that safe heat removal and the absence of surface boiling are provided. The maximum temperature of chromium disks in stationary mode does not exceed 122°C.

For the manufacture of the $^{51}$Cr neutrino source 26 disks of enriched chromium of two standard sizes (13 of each type, see section 3.3) were irradiated with a total mass of 4007.5 g. Figure 4 shows the IRD assembly.

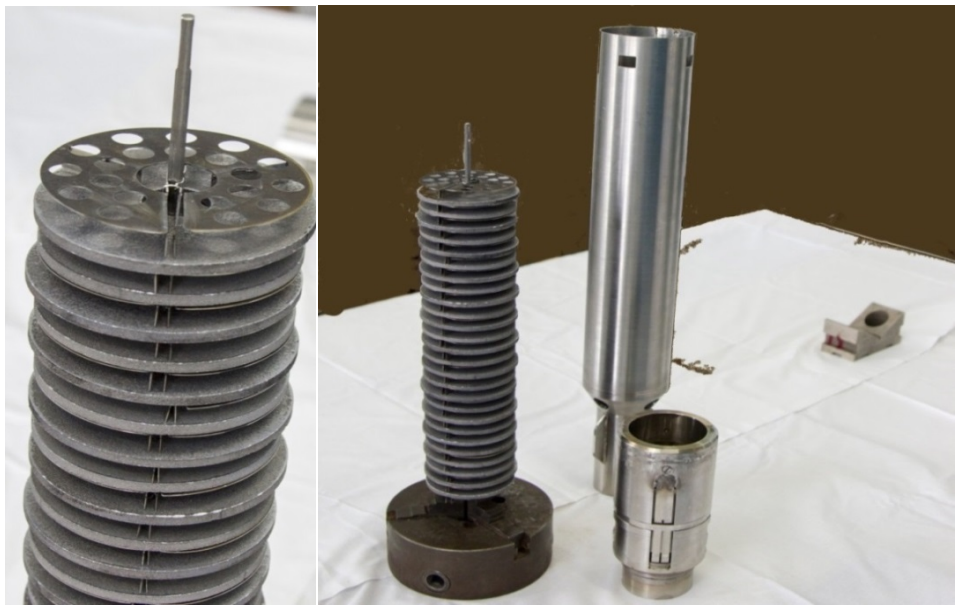

Figure 4. The irradiation device with chromium disks.

The disks were irradiated from 22 March 2019 to 2 July 2019 according to a specially developed schedule, the irradiation time was 75.4 effective days (Figure 5).

| March | | | | | | | | | | | | | | | | | | | | | | | | | | | | | | |
|---|---|---|---|---|---|---|---|---|---|---|---|---|---|---|---|---|---|---|---|---|---|---|---|---|---|---|---|---|---|---|
| 1 | 2 | 3 | 4 | 5 | 6 | 7 | 8 | 9 | 10 | 11 | 12 | 13 | 14 | 15 | 16 | 17 | 18 | 19 | 20 | 21 | 22 | 23 | 24 | 25 | 26 | 27 | 28 | 29 | 30 | 31 |

April

| 1 | 2 | 3 | 4 | 5 | 6 | 7 | 8 | 9 | 10 | 11 | 12 | 13 | 14 | 15 | 16 | 17 | 18 | 19 | 20 | 21 | 22 | 23 | 24 | 25 | 26 | 27 | 28 | 29 | 30 |
|---|---|---|---|---|---|---|---|---|---|---|---|---|---|---|---|---|---|---|---|---|---|---|---|---|---|---|---|---|---|

May

| 1 | 2 | 3 | 4 | 5 | 6 | 7 | 8 | 9 | 10 | 11 | 12 | 13 | 14 | 15 | 16 | 17 | 18 | 19 | 20 | 21 | 22 | 23 | 24 | 25 | 26 | 27 | 28 | 29 | 30 | 31 |
|---|---|---|---|---|---|---|---|---|---|---|---|---|---|---|---|---|---|---|---|---|---|---|---|---|---|---|---|---|---|---|

June

| 1 | 2 | 3 | 4 | 5 | 6 | 7 | 8 | 9 | 10 | 11 | 12 | 13 | 14 | 15 | 16 | 17 | 18 | 19 | 20 | 21 | 22 | 23 | 24 | 25 | 26 | 27 | 28 | 29 | 30 |
|---|---|---|---|---|---|---|---|---|---|---|---|---|---|---|---|---|---|---|---|---|---|---|---|---|---|---|---|---|---|

July

| 1 | 2 | 3 | 4 | 5 | 6 | 7 | 8 | 9 | 10 | 11 | 12 | 13 | 14 | 15 | 16 | 17 | 18 | 19 | 20 | 21 | 22 | 23 | 24 | 25 | 26 | 27 | 28 | 29 | 30 | 31 |
|---|---|---|---|---|---|---|---|---|---|---|---|---|---|---|---|---|---|---|---|---|---|---|---|---|---|---|---|---|---|---|

Figure 5. Calendar schedule of chromium irradiation in the SM-3 reactor (2019).

▉ Operation at a power of 90 MW with loaded chromium.
☐ Reactor shutdown for fuel assembly reload.

Figure 6 shows the dependence of the total activity of $^{51}$Cr on the irradiation time. According to the simulation the total activity of $^{51}$Cr was 3.55 MCi at the end of irradiation. The composition of chromium before irradiation and at the end of irradiation is shown in Table 6.

Figure 6. Dependence of the total activity of $^{51}$Cr source on the irradiation time (numerical simulations).

Table 6. Chromium composition before and after irradiation (numerical simulations).

| Date | Chromium mass, g | Isotope content, mass. % | | | | | Vanadium, g | $^{51}$V, %. |
|---|---|---|---|---|---|---|---|---|
| | | $^{50}$Cr | $^{51}$Cr | $^{52}$Cr | $^{53}$Cr | $^{54}$Cr | | |
| 22.03.2019 | 4007.5 | 96.55 | 0 | 3.2 | 0.2 | 0.05 | – | – |
| 02.07.2019 | 3939.5 | 95.50 | 0.973 | 3.270 | 0.20 | 0.058 | 69.0 | 100 |

## 5. Assembling and transportation the artificial neutrino source.

The neutrino source consists of 26 irradiated chromium disks of two standard sizes placed in a hermetic stainless-steel capsule (Figure 7a), a biological protection shield based on a tungsten alloy (W – 95%, Ni – 3%, Cu – 2%) with a thickness of ~ 30 mm and a weight of 42.8 kg, and a steel shell with a special cap for capturing the source by a manipulator (Figure 7b).

After irradiation, the IRD with disks was transferred to a shielding chamber for following operations: disassembly of the IRD, assembly of the source, capsule welding, capsule loading into biological protection and a transport container. All these operations were performed in about six hours.

When disassembling the device, there were no traces of chromium disc destruction, spalled spots, or the presence of chromium powder. The irradiated material was fully loaded into the source capsule, a 1 mm thick tungsten disk was loaded on the top of the column of chromium disks.

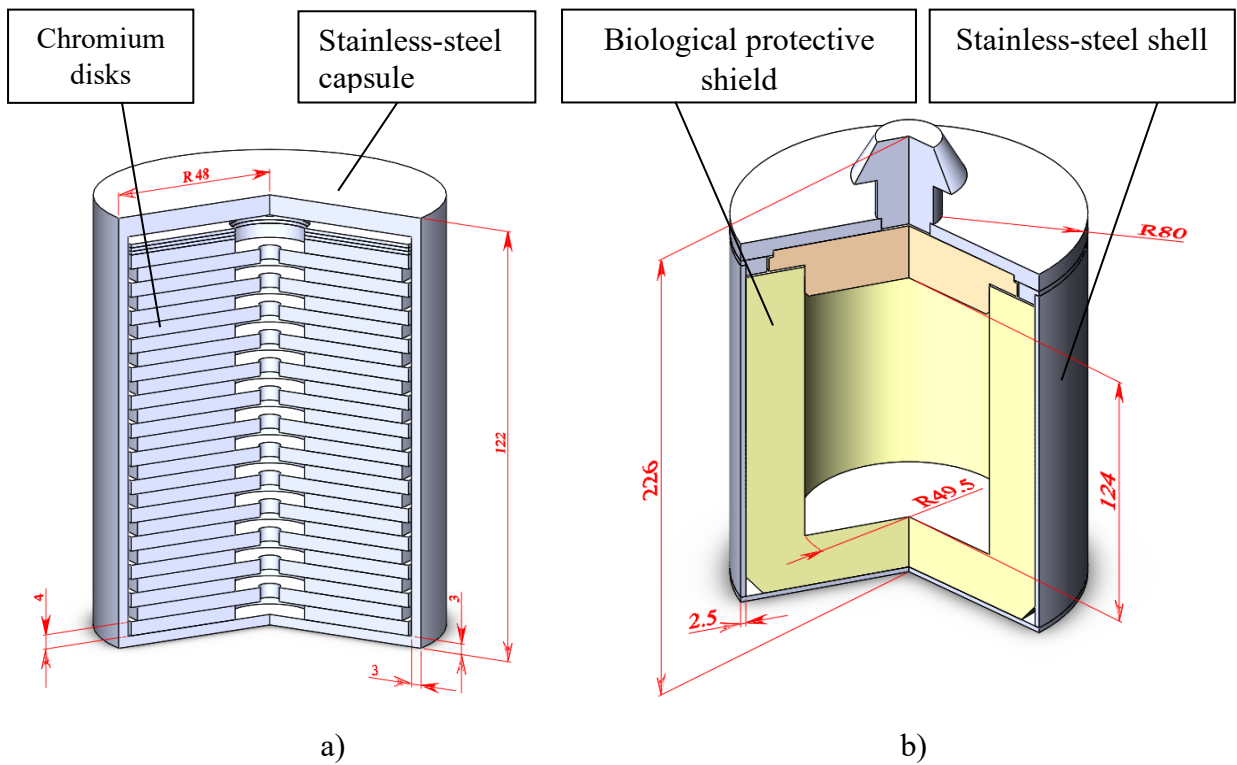

Figure 7. a) Stainless-steel capsule with chromium disks, b) biological protective shield made of tungsten alloy.

The key operation for remote assembly of the source in a shielding chamber is capsule sealing. The capsule was sealed by argon arc welding using a non-melting electrode on a specially designed equipment. Welding modes were developed on the source prototype considering the heating of the capsule to 350-450 °C. Visual and metallographic analysis of the prototype welds revealed no defects (cracks, pores, craters).

Biological protection shield with the neutrino source was installed in the protective container of the transport protective container (Figure 8). The radiation protection of the transport container consists of depleted uranium with a density of 18.5 g/cm$^3$, placed in a stainless-steel shell. To ensure the temperature conditions on the outer surface (the condition is not exceeding the temperature of 50°C on the outer surface), the transport container was installed in a protective container made of aluminum alloy AlMg3.

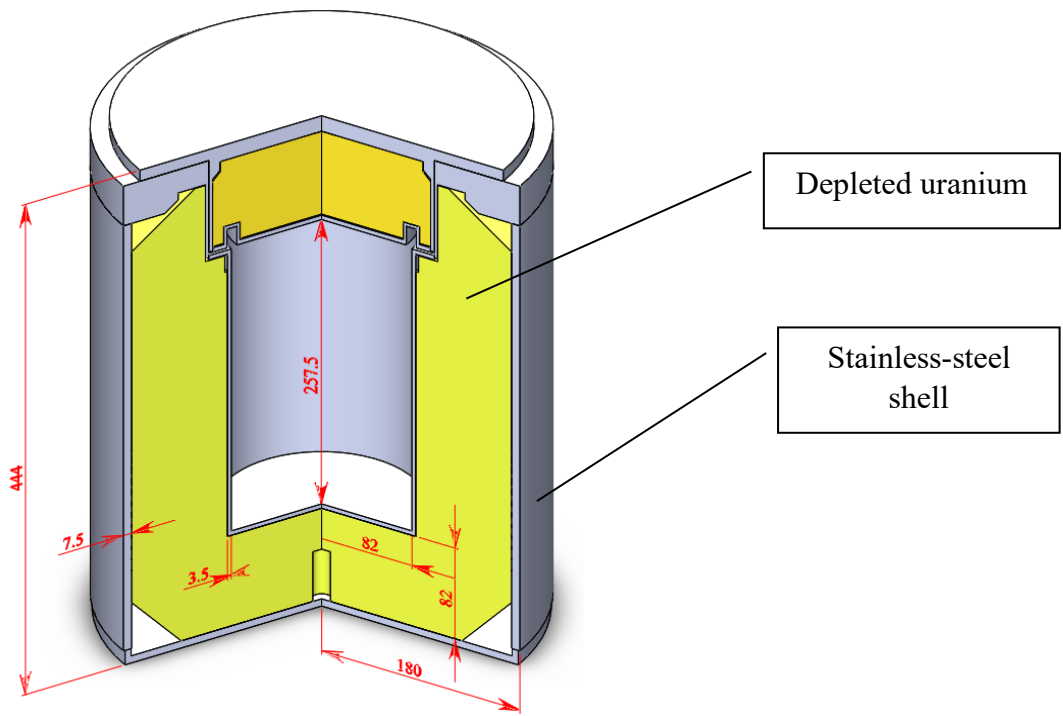

Figure 8. Transport protective container.

To confirm the possibility of safe transportation of the source and its operation in BNO, a calculated analysis of temperature conditions was carried out. The calculated analysis of thermal conditions was carried out using the Flow Simulation program of the SolidWorks package [15], considering thermogravitational convection and thermal radiation. Data on energy release, i.e. the amount of energy left behind from the x-rays and gammas interacting in the different layers of the source assembly, from Table 7 were used as initial parameters. When determining the energy release in the source, the MCNP program was used to calculate the transport of photons and electrons [9], the spectra of gamma radiation and auger electrons were taken from the ENDF/B-VIII nuclear data library [16].

Table 7. Energy release in the structural elements of the source (at the end of irradiation).

| Activity $^{51}$Cr, MCi | 3.55 |
|---|---|
| Chromium energy release, W | 525.2 |
| Energy release in the inner steel capsule of the neutrino source, W | 36.5 |
| Energy release in tungsten protection, W | 205.6 |
| Energy release in the transport protective container, W | <0.1 |
| Total energy release, W | 767.3 |

Based on the results of the calculation, the following parameters of the temperature field from the $^{51}$Cr neutrino source were obtained:

1. Source assembly (during manufacture):

* Maximum surface temperature - 412°C.

* Maximum temperature of chromium in the disks - 473°C.

2. A source installed in biological protection (for example, when operating in a BNO):

* Maximum surface temperature of biological protection with a source - 230°C.

* Maximum temperature of chromium in the disks - 408°C.

3. The source in biological protection is installed in a protective container, which in turn is placed in a protective container (for example, during transportation):

* The maximum temperature of the outer surface of the protective container - 49°C.

* Temperature of the protective container - 92°C.

* Biological protection temperature - 160°C.

* Maximum temperature of chromium in the disks - 387°C.

The delivery of the transport container with the source to BNO was carried out by special road transport. The time from reactor shutdown to delivery to the BNO (13:00 05.07.2019) was about 75 hours.

The neutrino source activity was measured by the calorimetric method [17]. At 14:02 05.07.2019 the activity of $^{51}$Cr was 3.414 ± 0.008 MCi in good agreement with the estimation. The measurements of photon spectra from the source and identification of lines from impurity radioactive elements in the source were performed to assess the contribution of their heat release to the total heat release of the source [17]. The contribution of radioactive impurities was about 5 × 10$^{-6}$ in the total heat release of the $^{51}$Cr source.

## Conclusions

As part of the BEST project to search for neutrino oscillations in a sterile state, an artificial neutrino source $^{51}$Cr was manufactured. The basis of the irradiation target was metal chromium enriched to 97% by $^{50}$Cr in the form of disks with a total mass of 4007.5 g. Chrome disks irradiated with thermal neutrons to an activity of > 3 MCi were assembled into a neutrino source with biological protection, which was delivered in a special transport container to the BNO for the BEST experiment. The activity of the source measured by the calorimetric method was amounted to 3.41 MCi at 13:00 05.07.2019. This is the most intense chemically pure neutrino source ever produced.

## Acknowledgments

We thank V.A. Rubakov for the constant interest and fruitful discussions and L.V. Kravchuk for great assistance in creating a neutrino source. This work was supported by Federal Agency for Scientific Organizations (FANO), Ministry of Education and Science of Russian Federation under agreement no. 14.619.21.0009 (unique project identifier no. RFMEFI61917X0009), and State Atomic Energy Corporation Rosatom.